\documentclass[10pt,conference]{IEEEtran}
\IEEEoverridecommandlockouts

\usepackage{cite}
\usepackage{amsmath,amssymb,amsfonts}
\usepackage{algorithmic}
\usepackage[linesnumbered,ruled]{algorithm2e}
\usepackage{amsthm}
\usepackage{array}
\usepackage{graphicx}
\usepackage{textcomp}
\usepackage{paralist}
\usepackage[caption=false,font=footnotesize]{subfig}
\usepackage{hyperref}
\usepackage{multirow}
\usepackage{url}
\usepackage{setspace}
\usepackage{tabu}
\usepackage{tablefootnote}
\usepackage{romannum}
\usepackage{enumitem}
\usepackage{arydshln}
\usepackage[normalem]{ulem}
\usepackage{threeparttable}
\usepackage{url}
\usepackage{xcolor}
\usepackage{xspace}
\usepackage{soul}
\usepackage{booktabs}
\usepackage{multirow}

\newcommand{\tool}{\textrm{AID}\xspace}
\newcommand{\company}{Huawei Cloud\xspace}
\begin{document}
\title{\tool: Efficient Prediction of Aggregated Intensity of Dependency in Large-scale Cloud Systems}


\author{
  \IEEEauthorblockN{
    Tianyi Yang\IEEEauthorrefmark{1},
    Jiacheng Shen\IEEEauthorrefmark{1},
    Yuxin Su\IEEEauthorrefmark{1}\thanks{Yuxin Su is the corresponding author.},
    Xiao Ling\IEEEauthorrefmark{2},
    Yongqiang Yang\IEEEauthorrefmark{2}, and
    Michael R. Lyu\IEEEauthorrefmark{1}
  }

  \IEEEauthorblockA{\IEEEauthorrefmark{1}Department of Computer Science and Engineering, The Chinese University of Hong Kong, Hong Kong, China.\\
    Email: \{tyyang, jcshen, yxsu, lyu\}@cse.cuhk.edu.hk}

  \IEEEauthorblockA{\IEEEauthorrefmark{2}Computing and Networking Innovation Lab, Cloud BU, Huawei\\
    Email: \{lingxiao1, yangyongqiang\}@huawei.com}
}


\maketitle

\begin{abstract}

%%%% background, cascading failures
Service reliability is one of the key challenges that cloud providers have to deal with.
In cloud systems, unplanned service failures may cause severe cascading impacts on their dependent services, deteriorating customer satisfaction.
Predicting the cascading impacts accurately and efficiently is critical to the operation and maintenance of cloud systems.
%%%% what current parctice do: tracing, no intensity
Existing approaches identify whether one service depends on another via distributed tracing but no prior work focused on discriminating to what extent the dependency between cloud services is.
%%%% the dep intensity problem, why it is important
In this paper, we survey the outages and the procedure for failure diagnosis in two cloud providers to motivate the definition of the intensity of dependency.
We define the intensity of dependency between two services as how much the status of the callee service influences the caller service.
%%%% what we propose
Then we propose \tool, the first approach to predict the intensity of dependencies between cloud services.
\tool first generates a set of candidate dependency pairs from the spans.
\tool then represents the status of each cloud service with a multivariate time series aggregated from the spans.
With the representation of services, \tool calculates the similarities between the statuses of the caller and the callee of each candidate pair.
Finally, \tool aggregates the similarities to produce a unified value as the intensity of the dependency.
%%%% the evaluation and benchmark
We evaluate \tool on the data collected from an open-source microservice benchmark and a cloud system in production.
The experimental results show that \tool can efficiently and accurately predict the intensity of dependencies.
%%%% applied in huawei
We further demonstrate the usefulness of our method in a large-scale commercial cloud system.

\end{abstract}


\section{Introduction}\label{sec:intro}

%%%%%% background: cloud computing, microservices
Service reliability is one of the key challenges that cloud providers have to deal with.
The common practice nowadays is developing and deploying small, independent, and loosely coupled cloud microservices that collectively serve users' requests.
The microservices that serve the same purpose are called cloud services\footnote{For simplicity, in this paper, ``cloud service'' and ``cloud microservice'' are interchangeable when they are used alone.}.
The microservices communicate with each other through well-defined APIs.
Such an architecture is called microservice architecture~\cite{microservice-architecture}.
The microservice architecture has been widely adopted in cloud systems because of its reliability and flexibility.
Under this architecture, microservice management frameworks like Kubernetes will be responsible for managing the life cycles of microservices.
Developers can focus on the application logic instead of the bothering tasks of resource management and failure recovery.

%%%%%% background, cascading failure in cloud systems
Although microservice management frameworks provide automatic mechanisms for failure recovery, unplanned service failures may still cause severe cascading effects.
For example, failures of critical services that provide basic request routing functions will impact the invocation of cloud services, slow down request processing, and deteriorate customer satisfaction.
Therefore, evaluating the impact of service failures rapidly and accurately is critical to the operation and maintenance of cloud systems.
Knowing the scope of the impact, reliability engineers can put more emphasis on services that have greater impacts on others.

% current parctice: tracing, no strength (introduce tracing)
A failed service will only affect services that will invoke it.
In other words, service invocations cause dependencies between services.
Many recent approaches~\cite{gmat,cloudscout} propose to use the dependencies of services to approximate their failure impact.
All the services and dependencies in a cloud system collectively construct a directed graph of services, which is also called a dependency graph.
Identifying whether one service depends on another in cloud systems can be well solved by industrial tracing frameworks like Dapper and Jaeger.
By using these frameworks, all the invocations between the caller and callee services can be recorded as traces that are composed of spans.
The attributes about each invocation, like duration, status, invoked service name, timestamp, etc., are recorded in each span.
Based on the spans, current dependency detection methods treat the dependency as a binary value indicating whether one service invokes another or not.

% the problem with current practice: ignores rich information in the traces

% the dep strength problem, why it is important
However, modeling the relations of services solely with binary dependencies is not precise enough.
To show the insufficiency of existing methods, we first conduct an empirical study on the outages of Amazon Web Service and \company.
We point out that it is inefficient to conduct failure diagnosis and recovery based on binary dependencies.
This is because the different dependencies of a cloud service impact the cloud service in different ways.
Manual examination of different dependencies without any priority is inefficient, especially in cloud systems where the number of dependencies could be large.
Based on this observation, we argue that it will be helpful if the dependency can be measured as a continuous value that indicates the intensity of this dependency.
Specifically, by checking services that are dependent on the failed service with large intensity values, on-call engineers (OCEs) can find the root cause of a system failure with a higher probability.
By recovering the services that are strongly dependent on the failed one, the whole system could be restored faster.

% However, no existing industrial tool or academic papers have worked on the strength of the dependencies between microservices.

% what we propose
To improve the reliability of cloud systems, in this paper, we propose \tool, an end-to-end approach to predict the intensity of dependencies between cloud microservices for cascading failure prediction.
We first generate a set of candidate dependency pairs from the spans.
Then we distribute each span into different fixed-length bins according to their timestamp and service name.
We calculate the statistics of all spans in each bin as the Key Performance Indicators (KPIs) for the bin.
The KPIs of one service form a multivariate time series that will be treated as the representation of the service's status.
For each candidate dependency pair, we calculate the similarities between the statuses of the two services in the pair.
Finally, we aggregate the similarities to produce a unified value as the intensity of the pair.

% the evaluation and benchmark
To show the effectiveness of \tool, we evaluate \tool on two datasets.
One is a simulated dataset, and the other is an industrial dataset.
For the simulated dataset, we deploy train-ticket, an open-source microservice benchmark system, simulate users' requests, and collect the traces.
For the industrial dataset, we collect the traces from a production cloud system.
Then we evaluate \tool on the datasets and compare its performance with several baselines.
The experimental results show that our proposed method can accurately measure the intensity of dependencies and outperform the baselines.
Furthermore, we showcase the successful usage of our method in a large-scale production cloud system.
In addition, we release both datasets to facilitate future studies.

% contributions
The main contributions of this work are highlighted as follows:
\begin{itemize}
  % \item We conduct a comprehensive industrial survey and an empirical study to identify the inefficiency of using binary-valued dependency for failure diagnosis and failure recovery.
  \item We propose \tool, the first method to quantify the intensity of dependencies between different services.
  \item The evaluation results show the effectiveness and efficiency of the proposed method.
  \item We release a simulated dataset and an industrial dataset from a production cloud system to facilitate future studies.
        % \item Additionally, Our method have been successfully applied in a leading public cloud provider, and helped greatly reduce manual maintenance effort.
\end{itemize}

\noindent \textbf{Organization.}
The remainder of this paper is organized as follows.
Section~\ref{sec:background} provides motivation and background knowledge that underpin our approach.
We describe our survey and empirical study on real outages that motivate the proposed method in Section~\ref{sec:motivation}.
Section~\ref{sec:approach} elaborates on the method in detail.
Section~\ref{sec:experiment} introduces the datasets, baselines and shows the experimental results.
Successful use cases of the proposed method in a production cloud system are demonstrated in Section~\ref{sec:case}.
We discuss the practical usage, the perceived limitations, and the possible threats to validity in Section~\ref{sec:discussion}.
Section~\ref{sec:relatedwork} introduces related works.
The last section, Section~\ref{sec:conclu}, concludes this paper and lists directions for future exploration.

\section{Background}
\label{sec:background}

In this section, we briefly describe the service-oriented architecture of cloud systems and the distributed tracing tools in cloud systems.
Then we present the main techniques, i.e., time series similarity analysis, that underpin our approach.

\subsection{The Architecture of Cloud Systems}
\label{sec:background:cloud}

%%%%%% The microservice architecture in the cloud, how it works
Modern cloud systems are often constructed from a complex and large-scale hierarchy of distributed software modules~\cite{berkeley-view-cloud}.
The common practice nowadays is to develop and deploy these software modules as cloud microservices that collectively comprise multiple large cloud services~\cite{aws-architecture}.
Microservices are small, independent, and loosely coupled software modules that can be deployed independently~\cite{microservice-architecture}.
Different microservices serve different responsibilities~\cite{DBLP:journals/internet/OppenheimerP02} like user authentication, resource allocation, virtual network management, billing, etc.
When an external request arrives at the cloud system, the request will be routed through the system and served by dozens of different cloud services and microservices.
The microservices communicate with each other through well-defined APIs and, therefore, can be refactored and scaled independently and dynamically to adapt to incidents like surges of requests and service failures~\cite{villamizar2015evaluating}.
Such an architecture is called microservice architecture~\cite{microservice-architecture}.

%%%%%% the pros and cons of the architecture
%%%%%% pros: scale, develop, etc
%%%%%% cons: hard to diagnose
The microservice architecture becomes increasingly popular due to its high flexibility, reusability, and scalability~\cite{DBLP:journals/software/BalalaieHJ16}.
It enables agile development and supports polyglot programming, i.e., microservices developed under different technical stacks can work together smoothly.
However, the loosely coupled nature of microservices makes it difficult for engineers to conduct system maintenance.
Different microservices in a large cloud system are usually developed and managed by separate teams.
Each team only has access to their own services as well as services that are closely related, which means they only have a local view of the whole system~\cite{wang2021fast}.

%%%%%% the fault model and fault tolerant mechanism
As a result, the failure diagnosis, fault localization, and performance debugging in a large cloud system become more complex than ever~\cite{DBLP:conf/asplos/GanZHCHPD19, DBLP:conf/ccgrid/WangXMLPWC18,DBLP:conf/infocom/ChenQZH14}.
Despite various fault tolerance mechanisms introduced by modern cloud systems, it is still possible for minor anomalies to magnify their impact and escalate into system outages.
As exemplified in Section~\ref{sec:motivation:survey}, when a cloud service enters an anomalous state and does not return results in a timely manner, other services that depend on it will also suffer from the increased request latency.
Such anomalous states can propagate through the service-calling structure and eventually affect the entire system, resulting in a degraded user experience or even a service outage.

\subsection{Distributed Tracing}
\label{sec:background:trace}

%%%%%% what tracing can do (summary)
For commercial cloud providers, it is crucial to troubleshoot and fix the failures in a timely manner because massive user applications may be affected even by a small service failure~\cite{DBLP:conf/icse/0003HLXZHGXDZ19}.
Distributed tracing is a crucial technique for gaining insight and observability to cloud systems.

% https://www.jaegertracing.io/docs/1.22/architecture/#span
\begin{figure}[t]
    \centering
    \includegraphics[width=0.9\columnwidth]{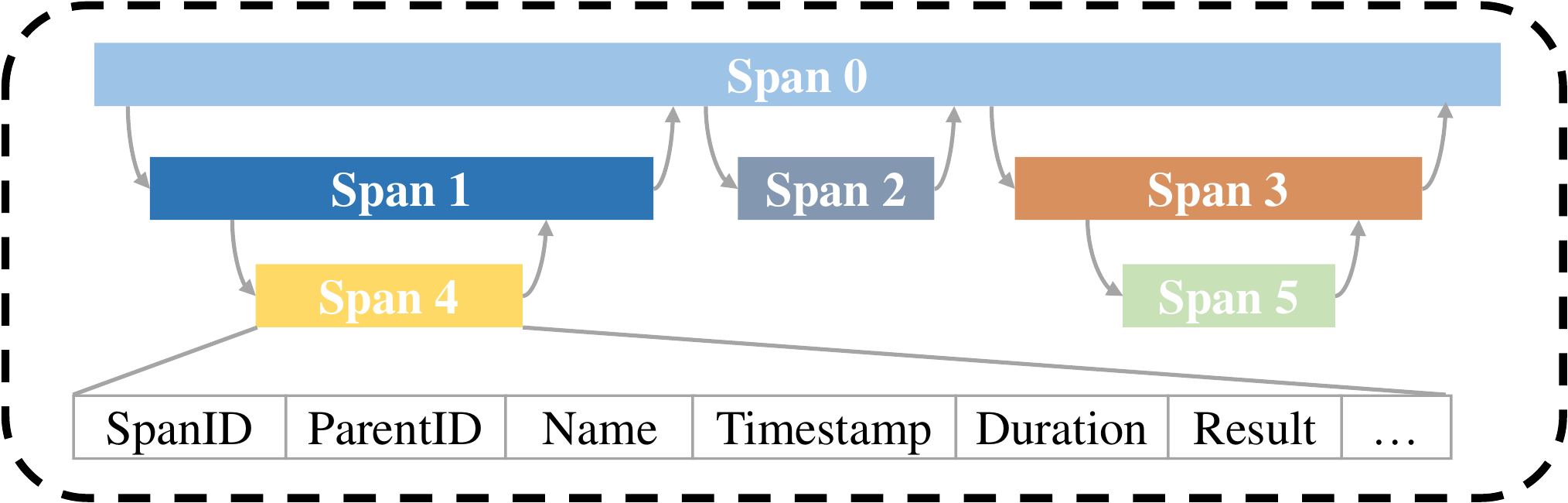}
    \caption{A trace log with six spans.}
    \label{fig:trace}
\end{figure}

%%%%%% what tracing can do (detail)
In large-scale cloud systems, a request is usually handled by multiple chained service invocations.
As clues to defective services are hidden in the intricate network of services, it is difficult for even knowledgeable OCEs to keep track of how a request is processed in the cloud system.
Distributed tracing provides an approach to monitor the execution path of each request.
%%%%%% what is tracing
For chained service invocations, e.g., service \texttt{A} invokes service \texttt{B}, and service \texttt{B} invokes service \texttt{C}, it is important to know the status of each service invocation, including the result, the duration of execution, etc.
By adding hooks to the services and microservices of the cloud system, a distributed tracing system~\cite{dapper, magpie, DBLP:conf/nsdi/FonsecaPKSS07} can record the contextual information of each service invocation.
Such records are called span logs, abbreviated as \texttt{spans}.
A span represents a logical unit of execution that is handled by a microservice in a cloud system.
All the spans that serve for the same request collectively form a directed graph of spans, as illustrated in Figure~\ref{fig:trace}.
Such a directed graph of spans generated by a request is called a piece of trace log, abbreviated as a \texttt{trace}.
A trace represents an execution path through the cloud system.
With a trace, engineers can track how the request propagates through the cloud system.
Collectively analyzing the traces of the entire cloud system can help engineers obtain in-depth latency reports that could assist failure diagnosis, fault localization, and surface performance degradation in the cloud system.

\begin{figure}[htbp]
    \centering
    \begin{tabular}{|l|l|}
        \hline
        Span ID            & e22f30bdbfd09134    \\
        \hline
        Parent Span ID     & b42a04bf18997d5d    \\
        \hline
        Name               & ts-preserve-service \\
        \hline
        Timestamp ($\mu$s) & 1618589098705000    \\
        \hline
        Duration ($\mu$s)  & 1126                \\
        \hline
        Result             & SUCCESS             \\
        \hline
        Trace ID           & c0d17d481f47bdd9    \\
        \hline
        Additional Logs    & ...                 \\
        \hline
    \end{tabular}
    \caption{A span generated by the train-ticket benchmark.}
    \label{fig:trace-sample}
\end{figure}

Although the actual implementation of distributed tracing systems varies a lot, the types of information they record are similar.
For clarity, we formally describe the attributes of spans as follows.
Suppose we have a trace $T$ composed of spans $\{s_1, s_2, ..., s_n\}$, a span $s_i \in T$ contains the following attributes\footnote{Other additional contextual information~\cite{opentracing-span} is omitted as we do not use them in our method.}:

\begin{itemize}
    \item $s_i^{id}$: The ID of span $s_i$,
    \item $s_i^{pid}$: The ID of the parent span of $s_i$,
    \item $s_i^{tid}$: The ID of the trace that $s_i$ belongs to,
    \item $s_i^{name}$: The name of service/microservice corresponding to $s_i$,
    \item $s_i^{ts}$: The time stamp of $s_i$,
    \item $s_i^{d}$: The duration of execution of $s_i$, and
    \item $s_i^{r}$: The result of execution of $s_i$.
\end{itemize}

Figure~\ref{fig:trace-sample} illustrates a span generated by the train-ticket benchmark~\cite{DBLP:journals/tse/ZhouPXSJLD21}.
It means that service \texttt{ts-preserve-service} was invoked at 04:58 on April 17, 2020.
The duration of execution is 1126 $\mu s$, and the execution result is SUCCESS.

\subsection{Time Series Similarity Analysis}
\label{sec:background:timeseries}

Time series data are ubiquitous.
One important task in time series data mining is to measure the similarity between two time series.
Similar to human intuition, the similarity measure is usually based on the similarity between the shapes of two time series~\cite{DBLP:conf/eit/FakhrazariV17}.
%%%%%% do we need this? Yes. Motivate the definition of DSW
% In particular, the similarity measure of time series should be consistent with human perception and intuition and abide the following properties~\cite{DBLP:reference/dmkdh/RatanamahatanaLGKVD10, DBLP:journals/csur/EslingA12}:
% \begin{itemize}
%     \item It should resemble human intuition and identify perceptually similar datasets, even if they are not identical mathematically.
%     \item It should not be restricted to particular time series or constraints.
%     \item It should be robust to noise, distortions and set of transformations like amplitude, scaling, temporal warping.
%     \item It should be able to capture global and local similarities.
% \end{itemize}

Dynamic time warping (DTW)~\cite{sakoe1978dynamic} is a widely-used similarity measure when two time series have the same overall component shapes but are not aligned on the timeline.
It attempts to align two time series along a timeline by distorting the timeline for one time series so that its converted form is better aligned with the second time series.
DTW was initially used in speech recognition applications~\cite{sakoe1978dynamic} and extended and optimized by many works~\cite{DBLP:conf/kdd/BerndtC94,DBLP:conf/mue/NiennattrakulR07,DBLP:conf/icdm/MueenHE14}.

\section{Motivations}
\label{sec:motivation}

The research described in this paper is motivated by the maintenance of a real-world cloud system in production.
In this section, we first survey thirteen publicly known service outages that severely affected Amazon Web Services (AWS) from 2011 to 2020.
Among the thirteen outages, we identify five that are related to service dependency and summarized the consequences of inappropriate management of service dependency.
Second, we empirically study the diagnosis records of five real outages in the cloud system of \company that are related to inappropriate management of service dependency.
% on the general practice of service failure diagnosis in the cloud system of \company.
Our study indicates that the information in the traces has not been used efficiently and current practice heavily relies on the engineers' familiarity with the dependencies in the system.
% Then we decompose the procedure of a service invocation and summarize the procedure of failure propagation incurred by service dependency.
Lastly, we propose to measure the intensity of dependency in terms of status propagation between dependent cloud services.
We demonstrate the usefulness of the intensity by motivating examples in real cloud systems.

\subsection{A Survey of the Outages in AWS}
\label{sec:motivation:survey}

\begin{table}[t]
  \centering
  \caption{Summary of AWS outages related to service dependency.}
  \label{tab:aws-summary}
  \begin{tabular}{@{}ccc@{}}
    \toprule
    \multirow{2}{*}{Date} & \multicolumn{2}{c}{Consequences}                 \\ \cmidrule(l){2-3}
                          & Cascading Failure                & Slow Recovery \\ \hline  %\cmidrule(r){1-1}
    Apr 21, 2011          & \checkmark                       &               \\
    June 29, 2012         &                                  & \checkmark    \\
    Oct 22, 2012          & \checkmark                       &               \\
    Aug 7, 2014           &                                  & \checkmark    \\
    Nov, 25 2020          & \checkmark                       & \checkmark    \\
    \bottomrule
  \end{tabular}
\end{table}

Service outages are inevitable in the cloud~\cite{DBLP:conf/sigsoft/ChenKLZZXZYSXDG20}.
In this section, we empirically analyzed over 1000 incidents of \company in 2019 and thirteen publicly known major outages\footnote{\url{https://aws.amazon.com/premiumsupport/technology/pes/}} of AWS from 2011 to 2020.
Among the incidents of \company, we found that improper service dependency is the most frequent reason for failures in \company.
Among the outage summaries of AWS, we also identified that five of the outages (38\%) are related\footnote{The outages are usually caused by various reasons that mutually affect each other. Service dependency is one of the reasons, so we use the word ``related''.} to service dependency.
As shown in Table~\ref{tab:aws-summary}, among the five outages that are related to service dependency, three of them are due to cascading failures triggered by erroneous upgrades of services.
During the failure recovery, the inappropriate dependencies lead to slow failure recovery in three outages.

%%%%%% Introduce AWS
AWS is the worldwide leading cloud provider.
It operates in many regions, each consisting of multiple Availability Zones (AZs).
Each AZ uses separate physical facilities and independently provides various cloud services~\cite{aws-architecture}, including Steam Data Processing (Kinesis), API Usage Analysis (Cognito), Customer Dashboard (Cloudwatch), Elastic Compute Cloud (EC2), Relational Database Service (RDS), Elastic Load Balancing (ELB), and Low-level Block Storage (EBS), etc.
For brevity's sake, we simplify the dependencies as 1) EC2, RDS, and ELB all depend on EBS, and 2) Cognito and Cloudwatch depend on Kinesis\footnote{The actual dependency relations between these services are complicated. We omit the details here.}.

%%%%%% Introduce failures
The outages on April 21, 2011, and October 22, 2012, are both caused by erroneous upgrades of EBS. When EBS failed, the services that depend on EBS, i.e., EC2,ELB, and RDS, are all affected. The cascading failures resulted in service disruptions of over 48 hours in the US-East-1 Region of AWS.

The outages on June 29, 2012, and August 7, 2014, are both triggered by the blackouts. After the blackout, the RDS and ELB services restarted quickly as expected, but they are still unable to fully recover because they both depend on EBS service which, at that time, can not recover simultaneously.
The slow failure recovery incurred by service dependencies affected the service availability for days in the US-East-1 Region and the EU West-1 Region of AWS.
As a follow-up optimization, ELB service reduced the dependency on EBS after the outage in 2014.

On November, 25 2020, the erroneous upgrade of Kinesis lead to its failure, cascadingly causing the failure of Cognito and Cloudwatch. More severely, during the recovery, AWS could not notify the customers via the normal way because the normal customer notification service also relied on Cognito. Due to the inner mechanism of Kinesis, the recovery of Kinesis took more than ten hours.
Thus the recoveries of Cognito and Cloudwatch were also slowed down.
As a follow-up optimization, Cognito and Cloudwatch services reduced the dependency on Kinesis after the severe outage.

\subsection{Drawbacks of Current Failure Diagnosis Methods}

\label{sec:motivation:procedure}

To gain more knowledge about the procedure of failure diagnosis in industrial circumstances, we first interviewed engineers in \company\footnote{AWS does not disclose the detailed procedures of failure diagnosis related to the five outages, so we cannot analyze the aforementioned outages in depth.}.
Then we summarize the procedure of failure diagnosis, and point out the drawbacks of current practice in \company.

%%%%%% what triggers the failure diagnosis
In \company, the failure diagnosis can be triggered by two systems, i.e., the customer support system and the monitoring system.
When a customer experiences a service disruption, the customer can submit a support ticket in the customer support system.
The on-call engineers will distribute the support ticket to the corresponding engineers responsible for the service.
The monitoring system, on the other hand, monitors the Key Performance Indicators (KPIs) and the logs of each service in the cloud system.
If the KPIs or the number of erroneous logs of one service increased abnormally or reached predefined thresholds, the monitoring system will send an alert to the corresponding engineers.
Upon receiving the support ticket or alert, engineers start diagnosing the failures.

%%%%%% how OCEs diagnose the service failures
% \jc{According to} the failure diagnosis records in \company, we identified the common practice for failure diagnosis.
We summarize the common practice of failure diagnosis in \company as follows.
Suppose the anomalous service is \texttt{A}, OCEs will first check whether the failure is caused by the faults of service \texttt{A} (e.g., an erroneous upgrade).
If so, the development team of service \texttt{A} will handle the failure.
If service \texttt{A} is in good condition, OCEs will analyze the status of all services that \texttt{A} depends on.
The status includes the number of calls, the error rate, etc.
If they found the failure of a service \texttt{B} is likely to cause the failure of service \texttt{A}, then engineers will continue to investigate service \texttt{B}.
Recall that all the services construct a directed graph where each node represents a service.
The failure diagnosis procedure can be viewed as a recursive search on the service dependency graph.

The practice works well in small cloud systems that contain tens of cloud services.
% However, as we will show in Section~\ref{sec:motivation:survey}, the service failure is usually caused by the failure of other services that \texttt{A} depends on.
However, the dependencies in large-scale cloud systems are much more complicated~\cite{DBLP:conf/asplos/GanZHCHPD19}, making manual failure diagnosis inefficient and difficult for engineers.
% According to our observation, one cloud service in \company depends on over \yi{} services on average and has multiple replicas.
% in large-scale cloud systems, one cloud service may have multiple replicas and depends on tens or more services.
Engineers may have trouble identifying the cause of the failure.
In this case, the development teams of all cloud services have to check whether the failure is caused by their corresponding services.
Sometimes engineers may infer the possible causes of a failure, but it heavily relies on the engineer's familiarity with the dependencies in the system.
In summary, the complex dependency relations in large-scale cloud systems make failure diagnosis difficult, and current practice is inefficient and dependent on the human experience.

\subsection{Intensity of Service Dependency}
\label{sec:motivation:measure}

\begin{figure}[htbp]
  \centering
  \includegraphics[width=0.95\columnwidth]{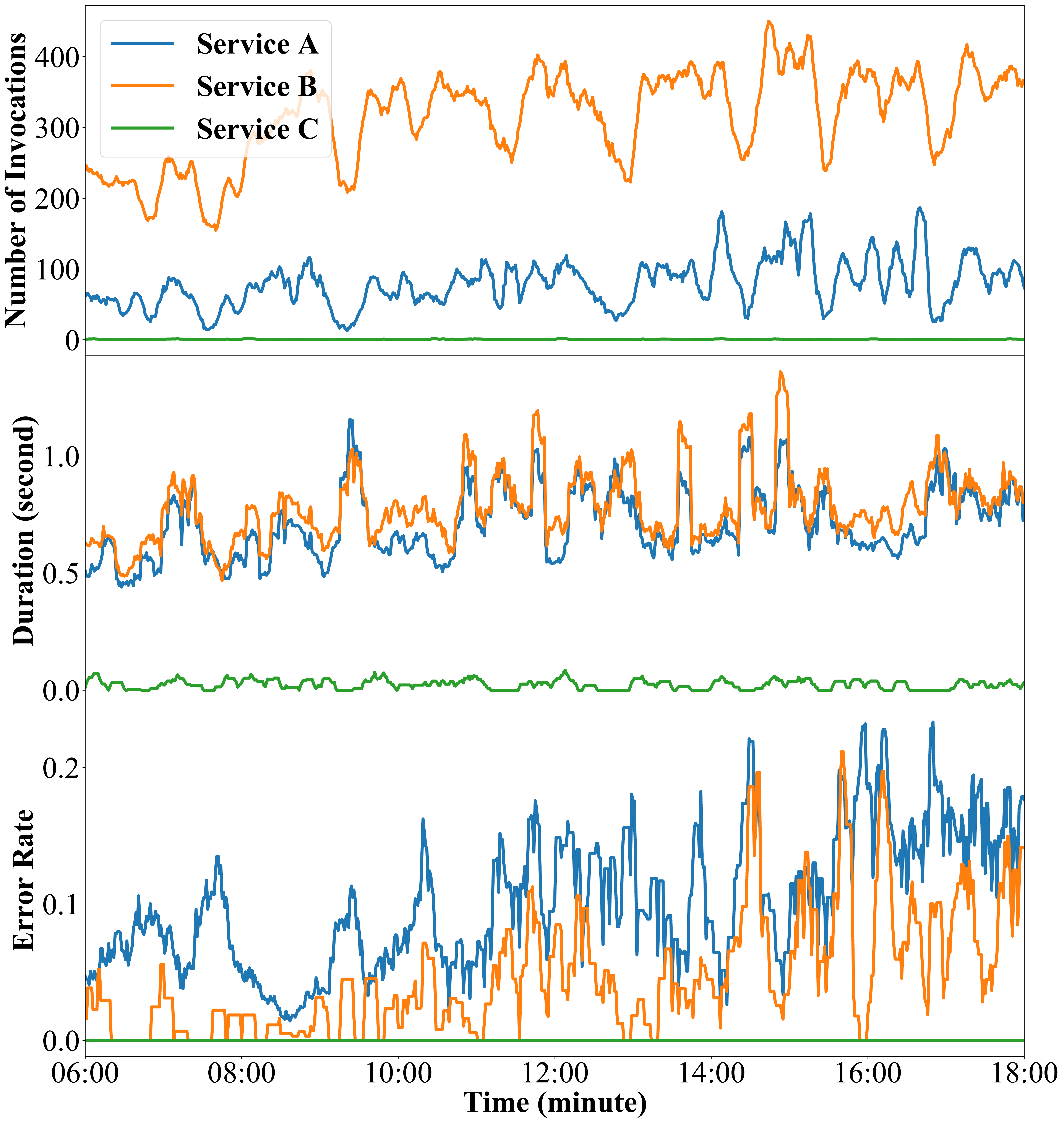}
  \caption{The statuses of service \texttt{A}, \texttt{B} and \texttt{C}. \texttt{A} invokes \texttt{B} and \texttt{C} but \texttt{B} has a greater effect on \texttt{A}. }
  \label{fig:motiv-all}
\end{figure}

A cloud system is composed of many services.
The dependency between two services is caused by one service invoking the other via predefined APIs.
Existing tools~\cite{miningdep,rippler,gmat} treat the dependency as a binary relation, i.e., if the caller service invokes the callee service, then the caller is dependent on the callee.
We suggest that this binary dependency metric is not fine-grained enough for cloud maintenance.
%%%%%% Observation of hw data.
Figure~\ref{fig:motiv-all} shows the statuses of three services\footnote{For confidentiality reasons, we cannot reveal the names of related services.} \texttt{A}, \texttt{B}, and \texttt{C} in \company.
Service \texttt{A} invokes both service \texttt{B} and service \texttt{C}.
Service \texttt{B} encountered failures.
The x-axis represents time in minute.
The y-axes represent the number of invocations per minute, the average duration of invocations per minute, and the error rate per minute of \texttt{A}, \texttt{B}, and \texttt{C}.
Although service \texttt{A} invokes service \texttt{B} and service \texttt{C}, it is obvious that the statuses of \texttt{B} and \texttt{C} influence the status of \texttt{A} in different degrees.
The reason is that the functionalities provided by service \texttt{A} and \texttt{B} are creating virtual machines, and allocating block storage, respectively.
Creating a virtual machine requires allocating one or more block storage.
Thus, the failure of service \texttt{B} inevitably affects service \texttt{A}.
On the contrary, due to the fault tolerance mechanism of service \texttt{A}, the failure of service \texttt{C} will not affect service \texttt{A} a lot.
Thus, it is more accurate to say that the intensity of dependency between service \texttt{A} and service \texttt{B} is higher than the intensity of dependency between service \texttt{A} and service \texttt{C}.
As can be seen in Figure~\ref{fig:motiv-all}, the similarity of the statuses reflect the difference in the intensities.

Ideally, if the development team of every cloud microservice accurately provides the intensity of dependencies for every dependent services, the failure diagnosis could be accelerated.
OCEs can prioritize the services that exhibit higher intensity of dependency instead of inspecting all the dependent services (Section~\ref{sec:motivation:procedure}) if they have accurate intensity information.
However, due to the complexity and the fast-evolving nature of cloud systems~\cite{DBLP:conf/globecom/AlamHZ18}, manually maintaining the dependency relations with intensity is very difficult.
As a result, OCEs often struggle in diagnosing failures due to the lack of intensities.
In order to relieve the pressure on OCEs, we propose to predict the intensity of dependency from the statuses of services.

\section{Approach}
\label{sec:approach}

%%%%%% overall figure
\begin{figure*}[htbp]
  \centering
  \includegraphics[width=1.5\columnwidth]{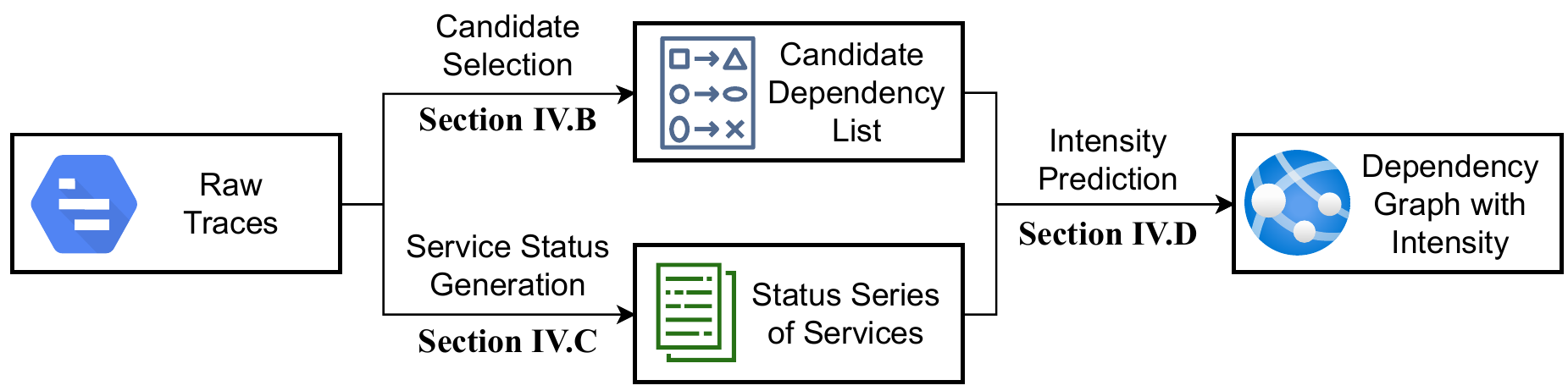}
  \caption{The overall workflow of \tool.}
  \label{fig:overview}
\end{figure*}

In this section, we present \tool, a framework for predicting the \underline{A}ggregated \underline{I}ntensity of service \underline{D}ependency in large-scale cloud systems.
We first present the overall workflow of \tool.
Then we elaborate on each step in detail, i.e., candidate selection, service status generation, and intensity prediction.

\subsection{Overview}
\label{sec:approach:overview}

%%%%%% talk about the ideas
The overall workflow of \tool is illustrated in Figure~\ref{fig:overview}.
\tool consists of three steps: candidate selection, status generation, and intensity prediction.
Given the raw traces, \tool first generates a set of candidate service pairs $(P,C)$ where service \texttt{P} directly invokes service \texttt{C} (Section~\ref{sec:approach:candgen}).
The intuition is that direct service invocation incurs direct dependency to some degree.
Indirect dependencies through the transitivity of service invocation will be discussed in Section~\ref{sec:discussion:usage}.
For status generation, we generate the status of all services~(Section~\ref{sec:approach:statusgen}).
The status of one service is composed of three aspects of dependency, i.e., number of invocations, duration of invocations, error of invocations.
Each aspect of the service's status contains one or more Key Performance Indicators (KPIs), depending on the actual implementation of the distributed tracing system.
A KPI is an aggregated value of a service status of all the spans of a service in a fixed time interval, e.g., 1 minute.
We use the statistical indicators of each aggregation as the values of the KPIs.
Motivated by the experience of engineers introduced in Section~\ref{sec:motivation:procedure}, we propose to predict the intensity of service dependencies from the similarity of the statuses of dependent services.
%%%%%% what is dep-intensity: the propagation
% \yi{The core idea of intensity prediction is to evaluate the degree of status propagation (e.g., number of calls, number of bugs, durations) of program status.}
The intuition behind using the similarity of time series is to evaluate the propagation of service statuses.
The intensity prediction step (Section~\ref{sec:approach:intensitypred}) predicts the intensity of dependency by measuring the similarity between two service's statuses.
The similarity between two service's statuses is a normalized and weighted average of the similarity of all the KPIs of the two services.
We calculate the similarity between two KPIs by a dynamic status warping algorithm.
Finally, \tool produces the dependency graph with intensity.

\subsection{Candidate Selection}
\label{sec:approach:candgen}

In general, direct service invocations can be divided into two categories, i.e., synchronous invocations and asynchronous invocations.
Modern tracing mechanisms can keep track of both synchronous and asynchronous invocations~\cite{opentracing-async}.
Given all the raw traces of the cloud system, in this step, we generate a candidate dependency set $Cand$.
The candidate dependency set $Cand$ contains service invocation pairs $(P_1,C_1), (P_2,C_2), \cdots, (P_n,C_n)$.
Each pair $(P_i,C_i)$ in the candidate dependency set denotes that the service named $P_i$ invokes the service named $C_i$ at least once.
Therefore, service $P_i$ depends on service $C_i$.
This step is to shrink the search space of possible dependent pairs because the service invocations indicate direct dependencies.

To generate the candidate dependency set, we need to know the name of the caller service and the callee service.
The name of callee service is clearly recorded in the span, but the name of the caller service is not.
Hence, we first augment each span $s$ by adding another attribute $s^{pname}$ which denotes the service name of the parent span.
Specifically, the augmentation of attribute $s^{pname}$ is achieved by 1) looking for another span $s'$ whose $id$ is the same as $s^{pid}$, and 2) set the $name$ of $s'$ as $s^{pname}$.
Then we iterate over all the spans and add $(s^{pname}, s^{name})$ to the candidate dependency set by the union operation.

For example, assuming the name of services are the same as the index of spans, the six spans in Figure \ref{fig:trace} will result in a candidate set of \{ ($Service_0$,$Service_1$), ($Service_1,Service_4$), ($Service_0$, $Service_2$), ($Service_0$,$Service_3$), ($Service_3$,$Service_5$) \} .

\subsection{Service Status Generation}
\label{sec:approach:statusgen}

In this step, we generate the status of all cloud services from the traces.
We start by defining the status of a cloud service (i.e., service status) and then describe the procedure of service status generation.

\subsubsection*{Definition of Service Status}
% Based on existing categorization of cloud failures~\cite{DBLP:conf/globecom/AlamHZ18}

A service invocation is composed of three logical components, i.e., the caller service, the callee service, and the network communication.
In particular, the caller service initiates an invocation to the callee service via the network.
The callee service then processes the invocation, during which it may invoke other services.
After the processing is finished, the callee service will send the result, e.g., status, to the caller service via the network.
Hence, we could derive three aspects of service invocations: \emph{initiation of invocation}, \emph{processing}, \emph{result}.
As service invocations occur repeatedly, the three aspects of service invocations can derive three aspects of service dependency:

%%%%%% three aspects of service status
\begin{itemize}
  \item \underline{\emph{Number of Invocations}}: The number of invocations from the caller to the callee,
  \item \underline{\emph{Duration of Invocations}}: The duration of invocations,
  \item \underline{\emph{Error of Invocations}}: The number of successful invocations from the caller to the callee.
\end{itemize}

\subsubsection*{Representation of Service Status}

In a cloud system, the spans record information about every invocation.
Intuitively, the status of a cloud service can be easily obtained from the spans of that service.
% To represent the status of a cloud service on a finer granularity,
Inspired by the common practice in cloud monitoring~\cite{aceto2013cloud}, we distribute the spans of one service into many bins according to the spans' timestamps.
Each bin accepts spans whose timestamp is in a short, fixed-length period.
We denote the length of the short period as $\tau$.
For example, the span shown in Figure~\ref{fig:trace-sample} will be put in the bin of \texttt{ts-preserve-service} at time 04:58, April 17 2020.
We can then represent the status of a cloud service in a short period by the statistical indicators of all the spans in the corresponding bin.

Formally, given all the spans in the cloud system over a long period $T$, we first initiate $\mathbf{S} \times \mathbf{N}$ empty bins of the predefined size $\tau$.
$\mathbf{S}$ is the number of microservices.
$\mathbf{N}$, determined by $T \over \tau$, is the number of bins.
Then we distribute all spans into different bins according to their timestamp $s^{ts}$ and service name $s^{name}$.
After that, we calculate the following three types of indicators as the KPIs for each bin.

\begin{itemize}
  \item $invo_{t}^{M}$: Total number of invocations (spans) in the bin;
  \item $err_{t}^{M}$: Error rate of the bin, i.e., the number of errors divided by the number of invocations;
  \item $dur_{t}^{M}$: Averaged duration of all spans in the bin;
        % \item $dstd_{t}^{M}$: Standard deviation of durations of all spans in the bin.
\end{itemize}

\noindent where $t$ is the time of the bin and $M$ is the microservice name of the bin.
If a service is not invoked in a particular bin (i.e., the corresponding bin is empty), all the KPIs will be zero.
In the end, we get the KPIs of every service $M$ at every period $t$.
Ordering the bins by $t$, we get three time series of KPIs for each cloud service, denoted as $invo^{M}$, $err^{M}$, and $dur^{M}$.
We name the time series of server KPIs as status series.

\subsection{Intensity Prediction}
\label{sec:approach:intensitypred}

\begin{algorithm}[t]
  \caption{Dynamic Status Warping}
  \label{algo:dsw}
  \LinesNumbered
  \KwIn{The status series of caller service and callee service $status^{P}, status^{C}$; duration series of callee $dur^{C}$, estimated round trip time $\delta_{rtt}$, max time drift $\delta_d$}
  \KwOut{The similarity between two status series}
  Set the warping window $w = \max (dur^{C}) + \delta_{rtt}$\\
  $K=length(status^{C})$ \\
  $N=length(status^{P})$ \\
  Initialize the cost matrix $\mathbf{C} \in \mathbb{R}^{K \times N}$, set the initial values as $+\infty$ \\
  $\mathbf{C}_{1,1} = (status^{P}_1 - status^{C}_1)^2$ \\

  \For(// Initialize the first column){$i=2 \ldots \min(\delta_d, K) $}{
    $\mathbf{C}_{i,1} = \mathbf{C}_{i-1,1} + (status^{P}_1 - status^{C}_{i})^2$
  }

  \For(// Initialize the first row){$j=2 \ldots \min(w + \delta_d, N) $}{
    $\mathbf{C}_{1,j} = \mathbf{C}_{1,j-1} + (status^{P}_j - status^{C}_{1})^2$
  }

  \For{$i=2 \ldots K$}{
    \For{$j=\max (2, i-\delta_d) \ldots \min (N, i +w+ \delta_d)$}{
      $\mathbf{C}_{i,j} = \min (\mathbf{C}_{i-1,j-1}, \mathbf{C}_{i-1,j}, \mathbf{C}_{i,j-1}) + (status^{P}_j - status^{C}_{i})^2$
    }
  }

  \Return{$\mathbf{C}_{K,N}$}
\end{algorithm}

In this paper, we define the intensity of dependency between two services as \emph{how much the status of the callee service influences the status of the caller service}.
The step of intensity prediction quantitatively predicts the intensity of dependency by measuring the similarity between two services' status series.
Specifically, we calculate the similarity of two different status series with dynamic status warping and aggregate all the similarities to get the overall similarity.

\subsubsection{Dynamic Status Warping}
Inspired by the dynamic time warping algorithm (DTW)~\cite{DBLP:conf/vldb/Keogh02}, we propose the dynamic status warping (DSW) algorithm (Algorithm~\ref{algo:dsw}) to calculate the distance between two status series.
DSW automatically warps the time in chronological order to make the two status series as similar as possible and get the similarity by summing the cost of warping.
It utilizes dynamic programming to calculate an optimal matching between two status series.
Given two services $P$, $C$, and their status series $invo^{P}$, $invo^{C}$, $err^{P}$, $err^{C}$, $dur^{P}$, and $dur^{C}$, the warping from the callee $C$ to the caller $P$ is specially designed for the cloud environment.
The design considerations include:

\noindent\emph{Directed warping:}
Due to the latency of the network and the time of processing, it takes some time for the status of the callee service to affect the status of the caller service.
Therefore, different from dynamic time warping, the time warping of DSW is directed, meaning that the matching from the callee to the caller can only happen in chronological order.

\noindent\emph{Adaptive propagation window:}
In cloud systems, after the round trip time ($\delta_{rtt}$) plus the duration of request processing, the caller can receive the result of an invocation.
Thus, the size of the directed warping window $w$ is automatically set as the maximum duration of the callee's spans plus $\delta_{rtt}$.

\noindent\emph{Time drift:}
The machine time may drift due to issues with time synchronization in cloud systems, so we add an undirected time drift $\delta_d$ to the warping window.

In summary, $status_i^{C}$ can only be matched with one of $[status_{i-\delta_d}^{P}, status_{i+w+\delta_d}^{P}]$.
The DSW returns the warping cost $\mathbf{C}_{M,N}$ as the measure of similarity.

% We use moving average with window size $w_{MA}$ to smoothen the data.
% In order to order both model the definite number and the fluctuation of each series, we conduct min-max normalization on each series.
% Both normalized and unnormalized series are used for the next step.

\subsubsection{Similarity Aggregation}

For all $(P_i, C_i) \in Cand$, we calculate similarities between their status series, denoted as $d_{invo}^{(P_i, C_i)}$, $d_{err}^{(P_i,C_i)}$, and $d_{dur}^{(P_i,C_i)}$.
We normalize the similarity across the whole candidate set with a min-max normalization with Equation~\ref{eq:norm}, where $status \in \{ invo, err, dur \} $.

\begin{equation}
  \label{eq:norm}
  \begin{aligned}
    d_{status}^{(P_i, C_i)} & = \frac{d_{status}^{(P_i, C_i)} - \min (d_{status}^{(P, C)})}{\max (d_{status}^{(P, C)}) - \min (d_{status}^{(P, C)})}
  \end{aligned}
\end{equation}

The intensity of dependency between $P_i$ and $C_i$ is the average similarity of all three similarities between their status series.

\begin{equation}
  \label{eq:agg1}
  I^{(P_i, C_i)} = \frac{1}{3}\sum_{status \in S}d_{status}^{(P_i, C_i)} \text{, } S=\{invo, err, dur\}
\end{equation}

Finally, we can build the dependency graph with intensity from the candidate set and the corresponding intensity values.

\section{Experiments}
\label{sec:experiment}

In this section, we evaluate \tool on both a simulated dataset and an industrial dataset.
Particularly, we aim to answer the following research questions (RQs):

\begin{itemize}
  \item \textbf{RQ1.} How effective is \tool in predicting the intensity of dependency?
  \item \textbf{RQ2.} What is the impact of different parameter settings?
  \item \textbf{RQ3.} What is the impact of different similarity measures?
  \item \textbf{RQ4.} How efficient is \tool?
\end{itemize}

\subsection{Experimental Setup}

\subsubsection{Dataset}

To show the practical effectiveness of \tool, we further conduct experiments on the simulated dataset and an industrial dataset from the cloud system of \company.
Since there are no existing datasets of trace logs, we deploy a benchmark microservice system to simulate a real cloud system.
We simulate user requests and collect the generated trace logs to construct the simulated dataset.
We release both datasets with the paper to facilitate future studies in this field\footnote{\url{https://github.com/OpsPAI/aid}}.

\begin{table}[t]
  \centering
  \caption{Dataset statistics.}
  \label{tab:dataset}
  \begin{tabular}{lll}
    \toprule
    % Dataset          & TT           & Industry \\ \midrule
    Dataset          & TT         & Industry\tablefootnote{We only labeled 75 dependencies that the engineers are familiar with.} \\ \midrule
    \# Microservices & 25         & 192                                                                                           \\ \midrule
    \# Spans         & 17,471,024 & About 1.0e10                                                                                  \\ \midrule
    % \# Spans         & 17,471,024 & About 1.8e13 \\ \midrule
    \# Strong        & 18         & 67                                                                                            \\ \midrule
    \# Weak          & 1          & 8                                                                                             \\
    \bottomrule
  \end{tabular}
\end{table}

\textbf{Simulated dataset}:
For the simulated dataset, we deploy train-ticket~\cite{DBLP:journals/tse/ZhouPXSJLD21}, an open-source microservice benchmark, for data collection.
Train-ticket is a web-based ticketing system with 25 microservices, through which users can search for tickets, reserve tickets, and pay for the reserved tickets.
An open-source tracing framework, Jaeger, is used to trace all the API calls.
To generate traces, we develop a request simulator that simulates normal users' access to the ticketing system.
The simulator will log in to the system, search for tickets, reserve a ticket according to the results of the search, and pay for the ticket.
Then we collect the traces from Jaeger and transform the traces into 17,471,024 spans.
The dataset is termed as ``TT'' in Table~\ref{tab:dataset}.

\textbf{Industrial dataset}:
%%%%%% 192 microservice, 80 services
Apart from the simulated dataset, we also collected traces from a region of \company to evaluate \tool.
To support tens of millions of users worldwide, the cloud system of \company contains numerous cloud services and microservices.
The service invocations are monitored and recorded by an independently developed distributed tracing system.
The complex dependency relations in the cloud system increase the burden of OCEs.
The OCEs can diagnose problematic microservices timely if the intensity of dependencies can be automatically detected in real-time.
To evaluate the practical effectiveness of our method, we collected a 7-day-long trace dataset with 192 microservices in April 2021.
The dataset is termed as ``Industry'' in Table~\ref{tab:dataset}.

\textbf{Manual labeling}:
Since our method is unsupervised, labels are only for evaluation.
Neither of the datasets has labels about the intensity of dependency, so manual labeling is needed.
We set two candidate labels for the intensity of dependency, i.e., ``strong'' and ``weak''.
Given a candidate dependency pair $(P,C)$, if the failure of service $C$ will cause the failure of service $P$, the intensity between $(P, C)$ should be labeled ``strong''; otherwise it should be labeled ``weak''.
For the simulated dataset, two Ph.D. students inspect the source code of all microservices and label every service dependency independently.
For the industrial dataset, several senior engineers are invited to manually label the intensity of dependency.
In both processes, disagreement on labels will be discussed until consensus is reached.
Finally, we convert the ``strong'' labels to $1$ and the ``weak'' labels to $0$ so that they can be effectively compared with the computed intensities.

The statistics of the datasets are listed in Table~\ref{tab:dataset}.
``\# Microservices'' denotes the number of microservices in the dataset.
``\# Spans'' denotes the number of spans in the dataset.
``\# Strong'' and ``\# Weak'' denote the number of dependencies that are labeled with ``strong'' or ``weak'' respectively.

\subsubsection{Baselines}
Since there is no existing work that measures the intensity of service dependency, we use Pearson correlation coefficient, Spearman correlation coefficient, and Kendall Rank correlation coefficient as the baseline.
Particularly, we calculate correlation on the status series of a candidate dependency pair $(P,C)$, denoted as $corrp_{status}^{(P,C)}$ and $corrs_{status}^{(P,C)}$.
For the baselines, we directly use the implementation from the Python package \texttt{scipy}.
We map the correlation to $[0,1]$ with the function $f(x) = (x+1)/2$.
The intensities of dependencies are then produced in the same way as Equation~\ref{eq:agg1}.

\subsubsection{Evaluation Metrics}

We employ Cross Entropy (CE), Mean Absolute Error (MAE), and Root Mean Squared Error (RMSE), as calculated in Equation~\ref{eq:metric}, to evaluate the effectiveness of \tool in predicting the intensity of dependency.

\begin{equation}
  \label{eq:metric}
  \begin{gathered}
    CE = \frac{1}{N} \sum_{i=1}^N - [ y_i \cdot \log (p_i) + (1 - y_i) \cdot \log (1-p_i) ] \\
    MAE = \frac{\sum_{i=1}^N |y_i - p_i| }{n} \\
    RMSE = \sqrt{\frac{\sum_{i=1}^N (y_i - p_i)^2}{N}}
  \end{gathered}
\end{equation}

Specifically, cross entropy calculates the difference between the probability distributions of the label and the prediction.
Mean absolute error and root mean squared error measures the absolute and squared error.
Lower CE, MAE, and RMSE values indicate a better prediction.

\subsubsection{Experimental Environments}
We run the experiments on the simulated dataset on a Linux server with Intel Xeon E5-2670 CPU @ 2.40GHZ and 128 GB RAM.
The experiments on the industrial dataset run on a Laptop with Intel Core i7 CPU @ 2.60 GHz and 16 GB RAM.

\subsection{RQ1: How effective is \tool in predicting the intensity of dependency?}

\begin{table}[]
  \centering
  \caption{Performance Comparison of Different Methods on Two Datasets}
  \label{tab:rq1}
  \begin{tabular}{@{}ccccc@{}}
    \toprule
    \multirow{2}{*}{Dataset}  & \multirow{2}{*}{Method} & \multicolumn{3}{c}{Metric}                                     \\ \cmidrule(l){3-5}
                              &                         & CE                         & MAE             & RMSE            \\ \midrule
    \multirow{4}{*}{TT}       & Pearson                 & 0.6872                     & \textbf{0.3305} & 0.4388          \\
                              & Spearman                & 0.7512                     & 0.3735          & 0.4697          \\
                              & Kendall                 & 0.6464                     & 0.3749          & 0.4577          \\
                              & \tool                   & \textbf{0.4562}            & 0.3435          & \textbf{0.3859} \\ \midrule
    \multirow{4}{*}{Industry} & Pearson                 & 0.6076                     & 0.4524          & 0.4563          \\
                              & Spearman                & 0.6030                     & 0.4501          & 0.4537          \\
                              & Kendall                 & 0.6258                     & 0.4636          & 0.4656          \\
                              & \tool                   & \textbf{0.3270}            & \textbf{0.1751} & \textbf{0.3044} \\ \bottomrule
  \end{tabular}
\end{table}

To study the effectiveness of \tool, we compare its performance with the baseline models on both the simulated dataset and the industrial dataset collected from \company.
For the parameters of \tool, we set the bin size $\tau = 1\ minute$, the estimated round trip time $\delta_{rtt} = 0$.
Specially, we set the max time drift $\delta_d = 1\ minute$ for the industrial dataset and set $\delta_d = 0$ for the simulated dataset.
We do this because the simulated dataset is deployed in a single server, so the time drift will not be a problem.
In addition, we use moving average to smoothen the status series for the baselines and our method.
The outputs are scalar values ranging from $0$ to $1$. A larger value indicates higher intensity.
The overall performance is shown in Table~\ref{tab:rq1}, where we mark the smallest loss for each loss metric and dataset.

\tool achieves the best performance on the industrial dataset and reduces the loss by 45.8\%, 61.1\%, and 33.2\% in terms of cross entropy, mean absolute error, and root mean squared error.
On the simulated dataset, \tool achieves the best performance in terms of cross entropy and root mean squared error.
Pearson correlation coefficient marginally outperforms \tool on the simulated dataset.
The improvement of \tool on the simulated dataset is smaller than that on the industrial dataset.
This is because the benchmark for simulation did incorporate very few fault tolerance mechanisms, making most of the dependencies strong.
Moreover, since the service invocations of the TT benchmark are very fast, the statuses of TT's services are relatively similar, making simple baselines and our approach perform similarly.

\subsection{RQ2: What is the impact of different parameter settings?}

\begin{figure}[htbp]
  \centering
  \includegraphics[width=0.95\columnwidth]{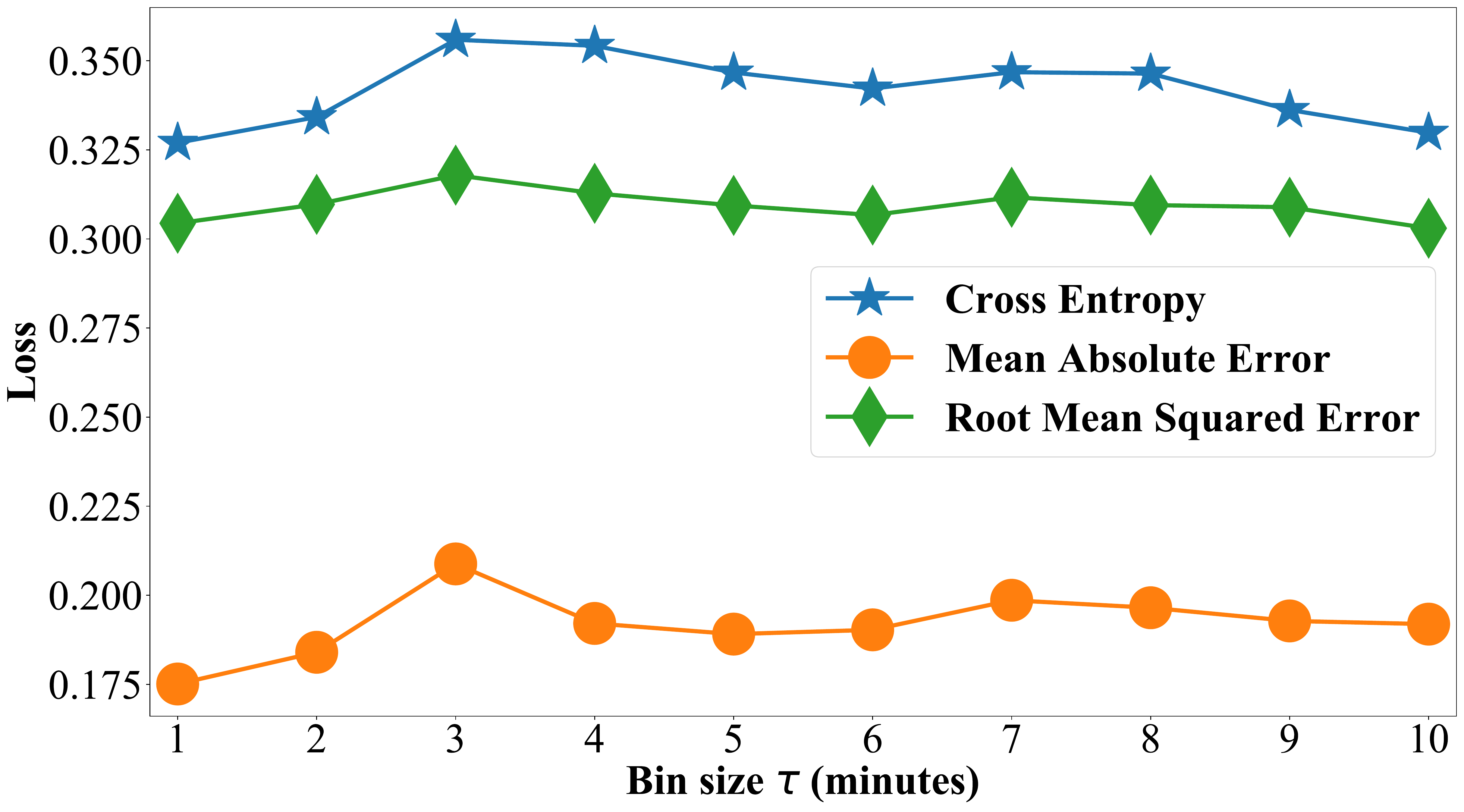}
  \caption{Prediction loss under different bin size $\tau$.}
  \label{fig:binsize}
\end{figure}

Since the estimated round trip time $\delta_{rtt}$ and the max time drift $\delta_d$ are minuscule, we only study the impact of the bin size $\tau$.
As the range of time of the simulated dataset is small, we only study the impact of the bin size $\tau$ in the industrial dataset.
In particular, we conduct experiments on with the bin size $\tau \in [1, 10] (minutes)$, and keep $\delta_{rtt} = 0$ and $\delta_d = 1\ minute$.
We did not set larger bin sizes because larger bin sizes result in more coarse-grained sampling of the service status, which will add difficulty to the similarity calculation in the subsequent DSW algorithm.

Figure~\ref{fig:binsize} shows the prediction loss under different bin size $\tau$.
The x-axis denotes the bin size and the y-axis shows the three loss metrics.
The results indicate that the impact of different bin sizes in a reasonable range is small, but $\tau = 1\ minute$ gives the best performance on the industrial dataset.

\subsection{RQ3: What is the impact of different similarity measures?}

\begin{table}[]
  \centering
  \caption{The impact of different similarity measures}
  \label{tab:rq3}
  \begin{tabular}{@{}ccccc@{}}
    \toprule
    \multirow{2}{*}{\begin{tabular}[c]{@{}c@{}}Dataset\\ /Bin size\end{tabular}} & \multirow{2}{*}{Method} & \multicolumn{3}{c}{Metric}                   \\ \cmidrule(l){3-5}
                                               &                         & CE                         & MAE    & RMSE   \\
    \midrule
    \multirow{2}{*}{\begin{tabular}[c]{@{}c@{}}TT\\ /1min\end{tabular}} & $\text{\tool}_{DSW}$    & 0.4562                     & 0.3435 & 0.3859 \\ \cmidrule(l){2-5}
                                               & $\text{\tool}_{DTW}$    & 0.4494                     & 0.3467 & 0.3832 \\
    \midrule
    \multirow{2}{*}{\begin{tabular}[c]{@{}c@{}}Industry\\ /1min\end{tabular}} & $\text{\tool}_{DSW}$    & 0.3270                     & 0.1751 & 0.3044 \\ \cmidrule(l){2-5}
                                               & $\text{\tool}_{DTW}$    & 0.3584                     & 0.1996 & 0.3169 \\
    \bottomrule
  \end{tabular}
\end{table}

We further study the impact of different similarity measures on both datasets.
$\text{\tool}_{DSW}$ denotes \tool that uses the proposed DSW to measure the similarity between status series.
$\text{\tool}_{DTW}$ denotes \tool that uses the DTW~\cite{DBLP:conf/vldb/Keogh02} to measure the similarity.
We keep the bin size $\tau = 1\ minute$ and the estimated round trip time $\delta_{rtt} = 0$ as usual.
Similar to previous experiments, we set the max time drift $\delta_d = 1\ minute$ for the industrial dataset and set $\delta_d = 0$ for the simulated dataset.

Table~\ref{tab:rq3} shows the performance of $\text{\tool}_{DSW}$ and $\text{\tool}_{DTW}$ on both datasets.
On the industrial dataset, the proposed DSW algorithm improves the performance, but on the simulated dataset, the performance is almost the same.
This is probably because the duration of spans in the simulated dataset is too small so that the effect of directed warping is weak.
The results imply that the proposed DSW algorithm works better in real-world cloud environments.

\subsection{RQ4: How efficient is \tool?}

The most time-consuming operations are the candidate selection and service status generation steps because we have to iterate over all the spans in the cloud system.
Theoretically, the time complexities of the candidate selection and service status generation steps are $O(S)$, where $S$ is the number of spans to process in the cloud system.
In practice, the industrial dataset contains about $1.0 \times 10^{10}$ spans, so we process it with a distributed computing service in \company.
Since the preprocessing is dynamically scheduled and mixed with other teams' tasks, we do not count the time spent on it.
For the intensity prediction step, the time complexity is $O(kN^2)$, where $N=\frac{T}{\tau}$ is the number of bins and $k$ is proportional to the warping window $w$.
In practice, the intensity prediction step takes $155$ seconds on average to process two status series both with $1440$ bins on a laptop.
Since the similarity calculation of different $(P,C)$ pairs are independent, we could easily parallelize the intensity prediction step to further improve the time efficiency.

\section{Case Study}
\label{sec:case}
%%%%%% how dep-intensity can help improve reliability

\begin{figure}[htbp]
  \centering
  \includegraphics[width=1\columnwidth]{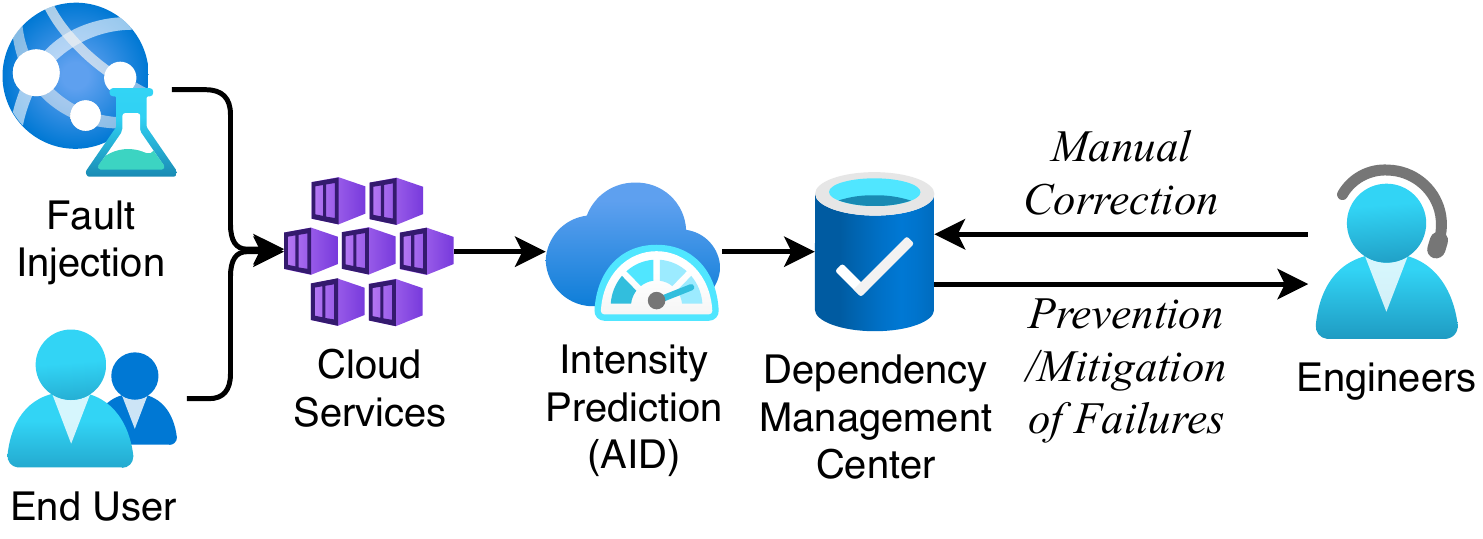}
  \caption{The use case of \tool.}
  \label{fig:case}
\end{figure}

In \company, \tool has been successfully incorporated into the dependency management system that serves hundreds to thousands of cloud services.
Figure~\ref{fig:case} illustrates the conceptual workflow.
\tool processes trace logs and continuously updates the aggregated intensity in the dependency management system.
The reliability engineers will categorize the intensity into different levels by referring to both the output of \tool and their domain expertise.
Then the dependency management system will provide reference to the engineers in optimizing dependencies and mitigating cascading failures.

\subsection{Optimization of Dependencies}

In a cloud system, service failures are inevitable, but we can prevent the failures from affecting other services by optimizing improper dependencies.
% \tool assists the prevention of cascading failures in two major ways.
% The first is the prediction of cascading failures.
% Exploiting the dependency graph with intensity, we can accurately predict the impact of failures caused by each service.
% The prediction is based on a propagation model designed for the cloud system of \company.
% As far as we know, OCEs will evaluate the impact of any possible failures before commencing each major upgrade of services.
\tool assists in the discovery of unnecessary strong dependency on critical cloud services.
If a critical cloud service depends on another service with high intensity, the dependency management system will remind the engineers to check whether the dependency is necessary.
If the dependency is unnecessary, the development team has to reduce the intensity of the dependency to improve the robustness of the critical cloud service.
Since \tool's deployment, more than ten unnecessary dependencies of critical cloud services have been discovered by \tool and optimized by the development team.

\subsection{Mitigation of Cascading Failures}
% failure diagnosis and recovery

\tool also assists in the mitigation of cascading failures.
During a cascading failure, \tool can provide the latest intensity of dependency to OCEs, so that they can diagnose service failures efficiently.
In addition, when a cascading failure occurs, OCEs can limit the traffic to critical cloud services and recover the dependencies marked as ``strong'' first.
By doing so, the service disruption can get under control.
Once a critical failure occurs, the manually confirmed ``strong'' dependencies will be treated with high priority.
We conduct field interviews with OCEs to collect feedback.
Based on the feedback, we have seen our method shedding light on reducing the impact of critical failures.

\section{Discussion}
\label{sec:discussion}

\subsection{Practical Usage and Perceived Limitations}
\label{sec:discussion:usage}

\subsubsection{Indirect Dependencies}

In this work, we mainly considered direct dependencies, which is caused by direct service invocations.
The proposed approach does not explicitly consider indirect dependencies through transitivity of service invocation because the intensity of indirect dependencies can be easily inferred from direct dependencies.
In practice, the intensity of indirect dependencies can be inferred by a ``cascading conduction mechanism'' that if \texttt{A} intensively depends on \texttt{B} and \texttt{B} intensively depends on \texttt{C} then \texttt{A} intensively depends on \texttt{C}.
The proposed approach also works well on dependencies caused by circuit breakers as long as the circuit breakers work transparently.

\subsubsection{Extension of Service Status}

In this paper, we only derive three aspects of service invocations, i.e., \emph{number of invocations}, \emph{duration of invocations}, \emph{error of invocations}.
We utilized them because they are part of the state-of-the-art tracing system.
Other aspects like the content of invocation responses can also be important to determine status.
In practice, cloud providers can incorporate additional information to extend the representation of service status in their own implementation of \tool.

\subsubsection{Limitations on Asynchronous Invocations}

Although modern tracing mechanisms can keep track of asynchronous invocations, \tool may suffer from inaccuracies when dealing with asynchronous invocations.
This is because the max time drift $\delta_d$ in Algorithm~\ref{algo:dsw} is hard to estimate for asynchronous service invocations.
Furthermore, if the traces of synchronous and asynchronous invocations are mixed, \tool may not work well since the time drift of synchronous and asynchronous invocations usually differs a lot.
We leave this problem as future work.

\subsection{Threat to Validity}
\label{sec:discussion:threat}

In this work, we identified the following major threats to validity.

%%%%%% adjust space
\vspace{0.25em}

\subsubsection{Labeling accuracy}

In this paper, we propose to measure the intensity of service dependency with \tool.
To evaluate the practical usage of \tool, we conduct experiments on a simulated dataset and an industrial dataset.
As it is a new relation between cloud services, manual labeling is needed for the evaluation.
The evaluation on the industrial dataset requires engineers to manually inspect the dependencies and label the intensity of dependencies.
Limited by the experience of engineers, the label may not be 100\% accurate.
The fast evolution of cloud services may also change their fault tolerance mechanism, resulting in inaccurate labels.
However, the engineers we invited have rich domain knowledge and are in charge of the architecture design of the cloud system of \company.
They also discuss with each other when there are disagreements.
Moreover, the labeled dependencies are the core cloud services in \company, so the intensity of dependencies are stable during the data collection period.
We believe the amount of inaccurate labels is small (if exists).
Most importantly, our method is unsupervised, so inaccurate labels will not affect the prediction results of the proposed method.

%%%%%% adjust space
\vspace{0.25em}

\subsubsection{Insufficiency of the simulation}

For the evaluation purpose, we deploy an open-source microservice benchmark to simulate a real cloud system.
The benchmark only contains 25 microservices, which is far below the number of cloud microservices in a real cloud.
Additionally, the implementation of the open-source benchmark did not fully consider the fault tolerance, resulting in only one weak dependency in the simulation.
Hence, the simulated dataset may not exhibit some common attributes of a real cloud system.
For example, the proportion of ``strong'' dependency in the simulated dataset is twice the proportion of ``strong'' dependency in the industrial dataset.
However, the insufficiency of the simulation will not hinder the practical usefulness of \tool in the real cloud system.
On the contrary, as we show in Section~\ref{sec:experiment}, the proposed method works better on the industrial dataset.
The experimental results on the simulated dataset only confirm the insufficiency of the simulation.

\section{Related Work}
\label{sec:relatedwork}

% \yi{cite}
% \cite{DBLP:conf/nsdi/ZhaiCPBTSZ20}
% \cite{DBLP:conf/icse/Zhao0PWWZCZNWWZ20}

\subsection{Cloud Monitoring}
\label{sec:relatedwork:monitor}

Monitoring cloud services properly with low overhead is the key to provide reliable services.
% tracing systems
Distributed Tracing, as a means of monitoring distributed cloud services, has been widely studied in the literature.
All the distributed tracing approaches can be classified as intrusive tracing and non-intrusive tracing.
Intrusive tracing requires modification to application code either in run time or at compile time.
Google proposes Dapper~\cite{dapper} to help engineers understand system behavior and reasoning about performance issues.
It reduces the tracing overhead by sampling and restricting the instrumentation number.
% Pivot Tracing~\cite{pivotsosp} provide the causal relationship of system events by combining dynamic instrumentation and happen-before join.
X-Trace~\cite{DBLP:conf/nsdi/FonsecaPKSS07} monitors and reconstructs the whole request path from a client by modifying all the network protocols and embedding the tracing data to the package header.

Non-intrusive tracing approaches do not require code modification and usually have a lower overhead.
Normally, these approaches leverage information like the system runtime logs and the source code to reconstruct the real event traces.
Zhao et. al.~\cite{lprof} propose lprof to reconstruct the execution flow of distributed systems using the runtime log of these systems.
lprof conducts static analysis on the call-graph of request processing code of the system to attribute a log output to a client request.
Chow et. al.~\cite{misty} also leverage system runtime logs to conduct performance monitoring and analysis.
They propose \textit{\"{U}berTrace} to reconstruct traces from the existing logs, then use \textit{The Mystery Machine} to construct a causal model and conduct analyses.
% Zhao et. al.~\cite{nonintru} propose Stitch, a non-intrusive performance monitoring tool, to obtain and present information that is helpful to locate performance bugs.
Stitch uses pattern matching on logs to reconstruct the hierarchical relationship of events in a system.
Pensieve~\cite{pensieve} automatically reconstructs a chain of causally dependent events that leads to a system failure exploiting the log files and system bytecode.

\subsection{Dependency Mining}
\label{sec:relatedwork:dependency}
Automatically discovering service dependencies is critical to cloud system administration and maintenance.
There are two major types of dependency mining approaches, i.e., passive dependency mining and active ones.
Passive dependency mining generates service dependency based purely on the runtime logs or KPIs.
Shah et. al.~\cite{dependencyanalysis} propose to use Recurrent Neural Networks (RNNs) to analyze and extract dependencies in KPIs and use the discovered dependencies to identify early indicators for a given performance metric, analyze lagged and temporal dependencies, and to improve forecast accuracy.
% \yi{causality related}
EIDefrawy et. al.~\cite{automated} use Transfer Entropy to passively mine the dependencies.
Luo et. al.~\cite{miningdep} apply log parsing and Bayesian decision theory to estimate the direction of dependencies among services.
They employ time delay consistency to reduce false dependencies.
Zand et. al.~\cite{know} construct a service correlation graph based on network measures and extract dependencies using hypothesis-testing.
They further compute an importance metric for network's components to facilitate administration.
CloudScout~\cite{cloudscout} employs Pearson Product-moment Correction Coefficient over machine-level KPIs such as TCP/UDP connection numbers and CPU utilization to calculate the similarity between different services.
The similarity measure is used to cluster different services together and to conduct VM consolidation based on the service clusters.
Unlike all these approaches that mostly use physical machine metrics to infer service dependencies, our method is designed for the emerging microservice architecture and utilizes the trace logs that directly record service invocations.

Active dependency mining requires modification to services.
Ma et. al. propose GMAT~\cite{gmat}, which generates service dependencies in the microservice architecture leveraging the reflection feature of Java and visualizes the dependencies to engineers.
Rippler~\cite{rippler} extracts the dependencies by randomly injecting temporal perturbation patterns in request arrival timings for different services and investigates the propagation of the patterns.
Wang et. al.~\cite{wang2019service} constructs a service knowledge graph using real-time measures, operational metadata, and business features.
They propose new metrics to measure the popularity of services based on their dependencies.
Novotny et. al.~\cite{novotny2015demand} focus on mining dependencies on the highly dynamic mobile networks.
They use local monitors to collect local views of dependencies and generate a global view of dependency on demand.

\section{Conclusion}
\label{sec:conclu}

In this paper, we first conduct a comprehensive empirical study on the maintenance of AWS and \company.
We identify the inefficiency in failure diagnosis and recovery with the binary-valued dependencies and define the intensity of dependency for the first time.
To facilitate cloud maintenance, we propose \tool, the first approach to predict the intensity of dependencies between cloud microservices.
\tool first generates a set of candidate dependency pairs from the spans.
\tool then represents the status of each cloud service with a multivariate time series aggregated from the spans and calculates the similarities between the statuses of the caller and callee of each candidate pair.
Finally, \tool aggregate the similarities to produce a unified value as the intensity of the dependency.
For the evaluation, we collect and manually label a new dataset from an open-source microservice benchmark and evaluate \tool on it.
Furthermore, we evaluate \tool using the data of \company and showcase the practical usage of \tool.
Both the evaluation results and case studies show the efficiency and effectiveness of \tool.
In the future, we plan to incorporate more information from the traces and other service logs for more accurate predictions.

% acknowledge fundings
\section*{Acknowledgment}

The work was supported by Key-Area Research and Development Program of Guangdong Province (No. 2020B010165002) and the Research Grants Council of the Hong Kong Special Administrative Region, China (CUHK 14210920).



\bibliographystyle{IEEEtran}
\bibliography{main}

\end{document}